\documentclass[12pt]{article}
\usepackage{multicol}
\usepackage{graphicx}
\usepackage{threeparttable}
\usepackage[round]{natbib}

\makeatletter
\renewcommand\section{\@startsection {section}{1}{\z@}%
                       {-1ex \@plus -0.25ex}%
                       {0.25ex}%
                       {\normalfont\large\bfseries}}
\renewcommand\subsection{\@startsection{subsection}{2}{\z@}%
                       {-1ex \@plus -0.25ex}%
                       {0.25ex}%
                       {\normalfont\bfseries}}
\makeatother

\newcommand\lesssim{\mathrel{\hbox{\rlap{\hbox{\lower4pt\hbox{$\sim$}}}\hbox{$<$}}}}
\newcommand{\Msun}{\mbox{M${}_\odot$}}%

\begin{document}

\begin{titlepage}
\thispagestyle{empty}


\begin{center}
{\Large\bfseries%
Gravitational Wave Astronomy Using Pulsars:\\
Massive Black Hole Mergers \& the Early Universe
}
\end{center}

\vspace*{1ex}

\begin{center}
{\large\bfseries
A White Paper for the Astronomy \& Astrophysics Decadal Survey}
\end{center}

\begin{center}
{\large\bfseries\centering
NANOGrav:\\
The North American Nanohertz Observatory for\\ Gravitational
Waves\\

\begin{center}
\includegraphics[width=1.0\textwidth]{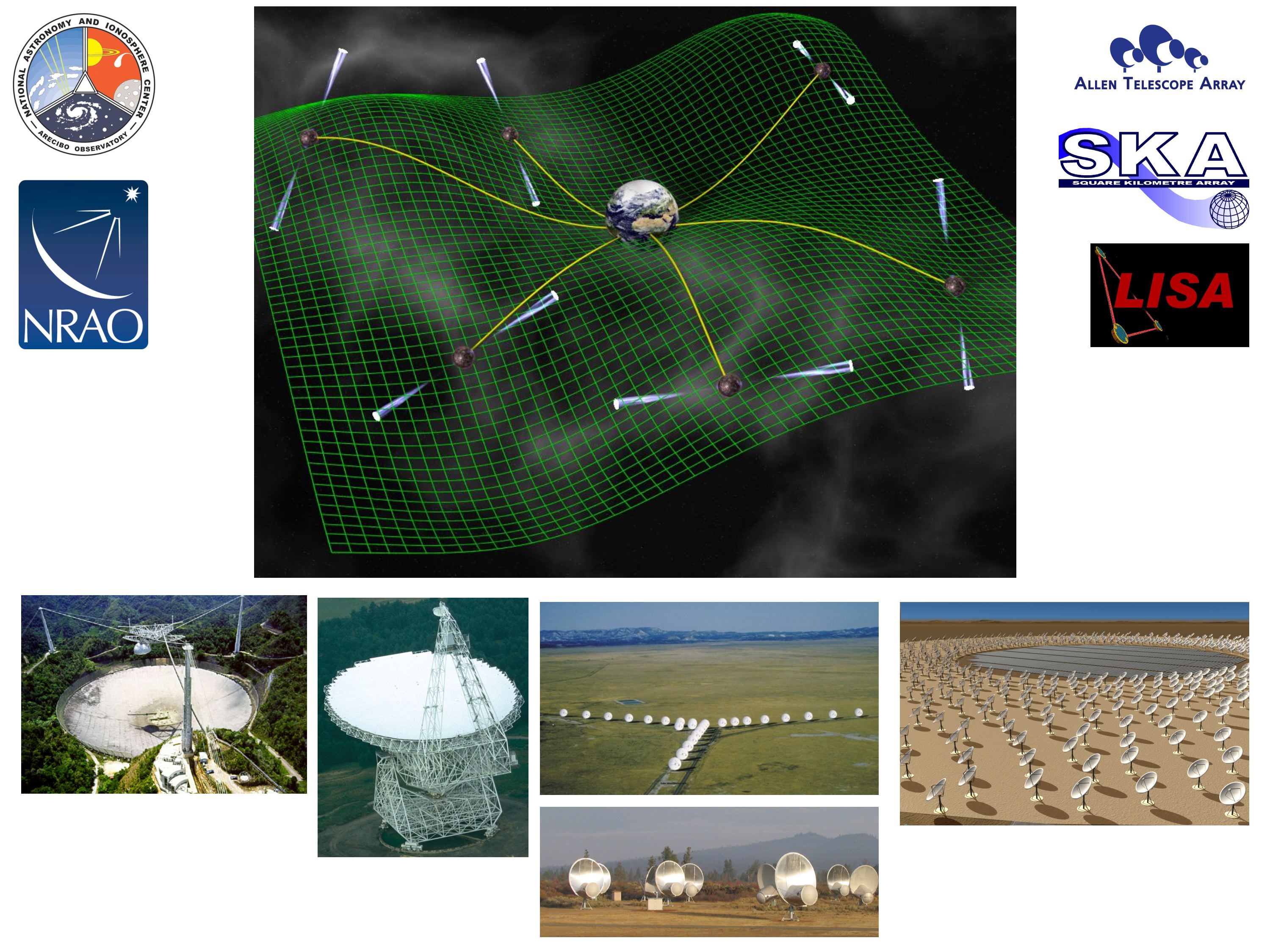}
\end{center}
}
\end{center}

{
\footnotesize
\noindent
{\bf Principal Authors:}
P.~Demorest (NRAO, 434-244-6838, \textit{pdemores@nrao.edu});
J.~Lazio (NRL, 202-404-6329, \textit{Joseph.Lazio@nrl.navy.mil});
A.~Lommen (Franklin \& Marshall, 717-291-4136, \textit{andrea.lommen@fandm.edu}) \\
{\bf NANOGrav Members and Contributors:}
A.~Archibald (McGill);
Z.~Arzoumanian (CRESST/USRA/NASA-GSFC);
D.~Backer (UC Berkeley);
J.~Cordes (Cornell);
P.~Demorest (NRAO);
R.~Ferdman (CNRS, France);
P.~Freire (NAIC);
M.~Gonzalez (UBC);
R.~Jenet (UTB/CGWA);
V.~Kaspi (McGill);
V.~Kondratiev (WVU);
J.~Lazio (NRL);
A.~Lommen (NANOGrav Chair, Franklin \& Marshall);
D.~Lorimer (WVU);
R.~Lynch (Virginia);
M.~McLaughlin (WVU);
D.~Nice (Bryn Mawr);
S.~Ransom (NRAO);
R.~Shannon (Cornell);
X.~Siemens (UW Milwaukee);
I.~Stairs (UBC);
D.~Stinebring (Oberlin)\\
{\bf This white paper is endorsed by:}
ATA;
LISA;
NAIC;
NRAO;
SKA;
US SKA;
D.~Reitze (LSC Spokesperson, U Fl.);
D.~Shoemaker (LIGO Lab, MIT);
S.~Whitcomb (LIGO Lab, Caltech);
R.~Weiss (LIGO Lab, MIT)
}

\vspace*{1ex}

\end{titlepage}

\section{Science Opportunity: Exploring the Low-Frequency Gravitational
	Wave Spectrum}\label{sec:opportunity}

Gravitational waves are fluctuations in the fabric of spacetime
predicted by Einstein's theory of general relativity.  Using a
collection of millisecond pulsars as high-precision clocks, the nHz band
of this radiation is likely to be detected within the next decade
\citep{Jenet05detect}.  The fundamental questions that will be addressed
by these studies are:

\vspace*{-1.5ex}
\begin{enumerate}

\item
{\bf What is the nature of space and time?} We suspect the local
spacetime metric is perturbed by the cumulative effect of gravitational
waves (GWs) emitted by numerous massive black hole (MBH) binaries.  What
is the energy density contained in this stochastic background of GWs?
\vspace*{-0.5ex}

\item
{\bf How did structure form in the Universe?} Detection of GWs in the
pulsar timing band will tell us whether MBHs formed through accretion
and/or merger events.
\vspace*{-0.5ex}

\item
{\bf What is the structure of individual MBH binary systems?}
Recovering the gravitational waveform from individual systems will give
us unprecedented insight.
\vspace*{-0.5ex}

\item
{\bf What contribution do cosmic strings make to the GW background
(GWB)?}  The detection of cosmic strings would open a window into the
early universe at a time inaccessible via the electromagnetic spectrum.
\vspace*{-0.5ex}

\item
{\bf What currently unknown sources of GW exist in the Universe?} Every
time a new piece of the electromagnetic spectrum has been opened up to
observations (e.g. radio, X-rays, and $\gamma$-rays), new and entirely
unexpected classes of objects have been discovered.  
\vspace*{-0.5ex}

\end{enumerate}
\vspace*{-1.5ex}

The existence of GWs has already been inferred via the Nobel
Prize-winning observations of the orbital decay of the PSR~B1913+16
binary system \citep{Hulse75}.  While compelling and entirely consistent
with general relativity, the behavior of this system offers only
indirect evidence for GWs -- the objective for the 21${}^{\mathrm{st}}$
century is the \emph{direct} detection and exploitation of GWs as a
non-photonic probe of the Universe.

Millisecond pulsars are old neutron stars that have been spun-up by mass
accretion from a companion star to spin rates of hundreds of Hz.  The
rotational stability of these pulsars surpasses the majority of
``normal'' pulsars, and rivals that of atomic clocks.  Pulsars emit a
beam of radio waves that sweeps past the Earth once per rotation,
appearing to us as a series of pulses.  By precisely measuring the times
of arrival of the the radio pulses on Earth, we can search for tiny
perturbations due to GWs.  This is distinct from detecting GWs emitted
by the pulsars themselves.  Rather, pulsar timing provides a means to
detect any gravitational radiation crossing the Earth--pulsar line of
sight, potentially from sources far outside our galaxy.

All modern GW observatories operate on the principle that passing GWs
cause tiny deviations in the distance between point masses.  GWs are
detected as changes in the light travel time between the points.  In the
case of the \emph{pulsar timing array} (PTA) the path from each pulsar
to Earth forms an arm of the GW detector.  This detector is most
sensitive to GWs with periods comparable to the total observation
timespan, typically 1--10 years, which corresponds to nHz frequencies.
The most desirable PTA involves millisecond pulsars evenly distributed
on the sky.  A passing GW modifies the spacetime around the Earth in a
manner that produces correlated shifts in the pulse times of arrival
from the different pulsars.

The North American Nanohertz Observatory for Gravitational Waves
(NANOGrav)\footnote{\texttt{http://www.nanograv.org}} is an organization
of astronomers, primarily from the U.S.  and Canada, working to achieve
GW detection using pulsar timing.  Current projects include ongoing
high-precision pulsar timing programs at Arecibo
Observatory\footnote{The Arecibo Observatory is a facility of the
National Astronomy and Ionosphere Center, operated by Cornell University
under a cooperative agreement with the National Science Foundation} and
the Green Bank Telescope (GBT)\footnote{The National Radio Astronomy
Observatory is a facility of the National Science Foundation operated
under cooperative agreement by Associated Universities, Inc}.  Here we
describe the potential sources of low-frequency GWs
(\S\ref{sec:context}) and the current status and key advances needed for
the detection and exploitation of GWs through pulsar timing
(\S\ref{sec:advance}).

\section{Science Context: Gravitational Wave Astrophysics}
\label{sec:context}

\subsection{Mergers of MBH Binaries}\label{sec:merger}

It is now well established that mergers are an essential part of galaxy
formation and evolution, and that massive black holes (MBHs,
$M>10^6\,\Msun$) exist in the nuclei of most, if not all, large galaxies
\citep[see e.g.][]{Ferrarese00}.  Consequently, the product galaxy of
many mergers will contain two MBHs.  Due to dynamical friction, these
two MBHs sink toward the center of the resulting galaxy's potential.  

As a MBH binary hardens, the strength of its GW emission increases.
Once the system reaches a point where its semi-major axis is
$\lesssim$1~pc, GW emission becomes the dominant form of energy loss,
and the two MBHs continue to spiral towards each other.  Further, as the
binary tightens, the frequency of the GW emission increases.  For
reference, an MBH binary with total mass $M$ and semi-major axis $a$
produces GWs with a frequency
\begin{equation}
f \sim 1\,\mathrm{nHz}\left(\frac{M}{10^9\,\Msun}\right)^{1/2} 
  \left(\frac{a}{1000\,\mathrm{AU}}\right)^{3/2}.
\label{eqn:f}
\end{equation}

Despite theoretical difficulties in fully understanding the production
of these hard binary systems, observational results continue to provide
evidence for their existence.  One dramatic piece of evidence is the
galaxy B0402$+$679, where high resolution radio imaging reveals two
radio-loud nuclei separated by only 7~pc \citep{rtzppr06}.  Follow-up
observations of SDSS galaxies show strong evidence for MBH binaries with
semi-major axes less than 1~kpc.  A small number of these galaxies show
[O\,\textsc{iii}] emission offset from the systemic redshift of the host
galaxy.  In two specific cases double-lined [O\,\textsc{iii}] profiles
have been found, suggesting double AGN \citep{cgn+08}.  Most recently, a
dual-broad-line QSO system has been identified where the inferred MBH
binary separation is only 0.1~pc \citep{boroson09}.

\begin{figure}[t]
\begin{center}
\includegraphics{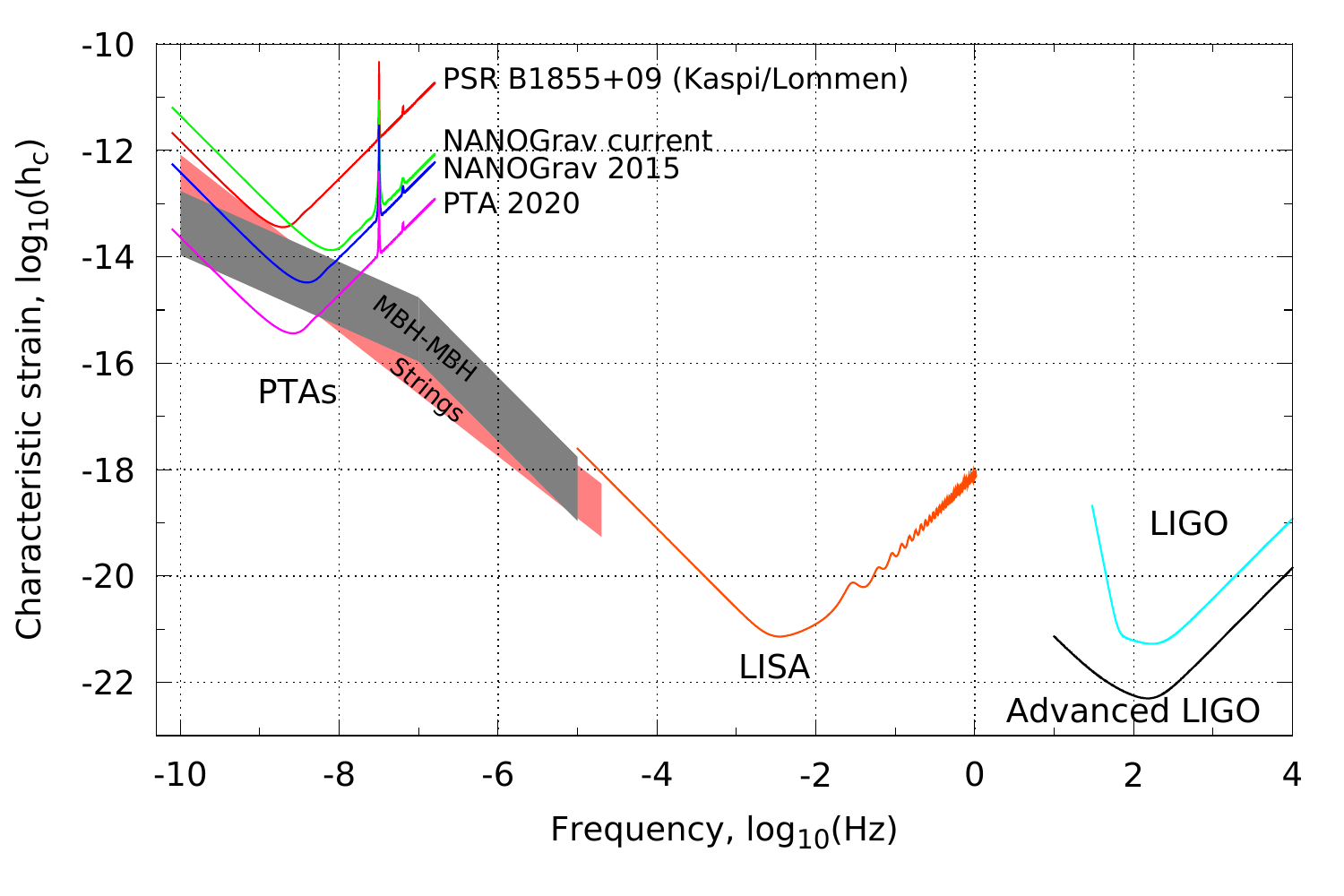}
\end{center}
\vspace*{-7ex}
\caption{\label{fig:gwb} Comparison of current and planned GW detectors,
showing characteristic strain ($h_c$) sensitivity versus frequency along
with expected source strengths.  The Laser Interferometer Gravitational
Wave Observatory (LIGO), the Laser Interferometer Space Antenna (LISA)
and PTAs occupy complementary parts of the GW spectrum.}
\vspace*{-1ex}
\end{figure}

\begin{figure}[t]
\begin{center}
\includegraphics{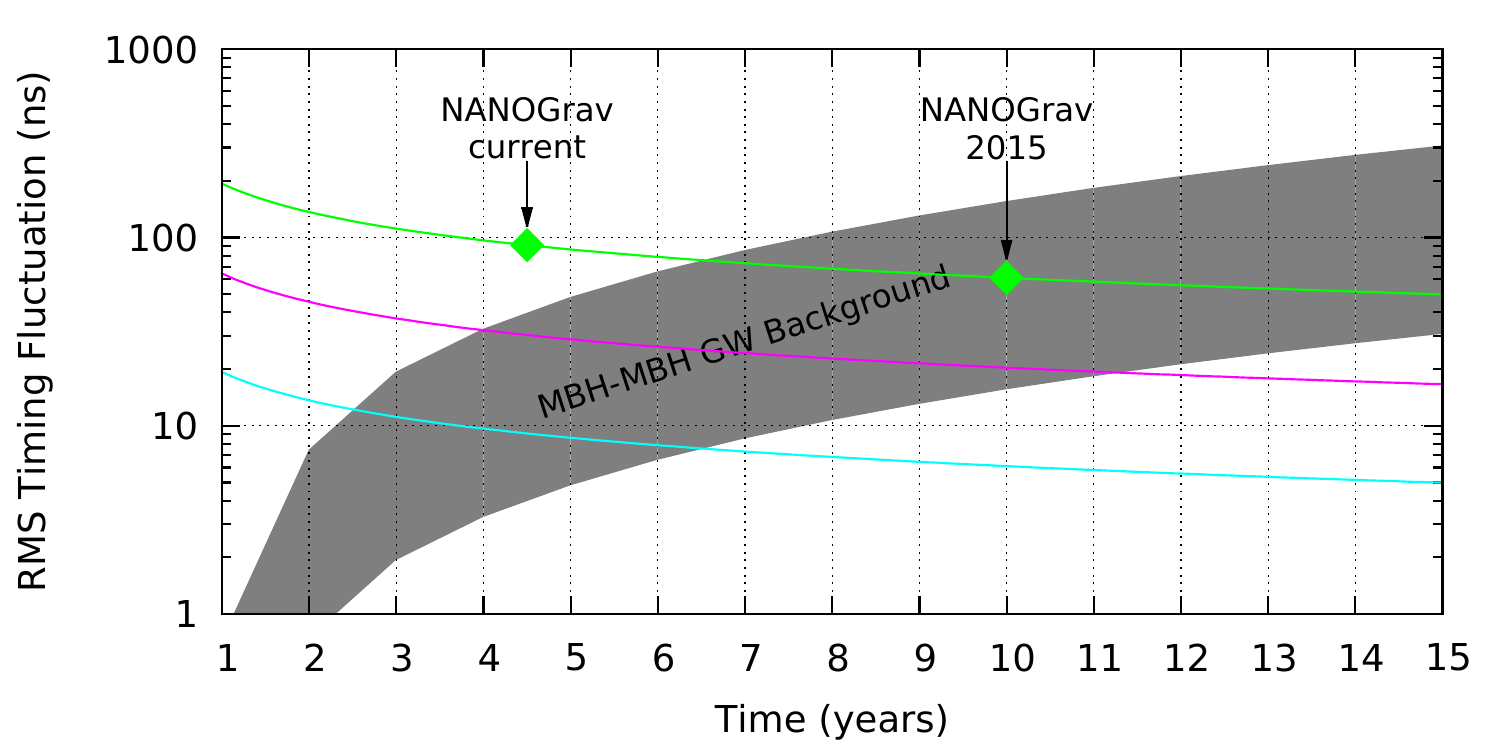}
\end{center}
\vspace*{-6ex}
\caption{\label{fig:sensitivity}%
PTA sensitivity versus times for several scenarios: The current NANOGrav
observing program, and potential future PTAs of 3 and 10 times better GW
sensitivity.  The shaded area shows the expected amplitude range of the
MBH-MBH GW signal (see \S\ref{sec:merger}).
}
\vspace*{-1ex}
\end{figure}

The general relativistic metric perturbation amplitude due to a GW is
commonly presented in terms of a dimensionless quantity called the
\emph{characteristic strain}~$h_c$.  The ensemble of MBH binaries is
expected to produce a GWB whose amplitude spectrum has a power-law
shape, $h_c(f) \propto f^\alpha$, for a GW frequency~$f$, where it is
predicted that $\alpha = -2/3$ \citep{Phinney01, Jaffe03}.  The strain
amplitude is less certain, but expected to be in the range $10^{-16}$ to
$10^{-15}$ at $f=1$~yr$^{-1}$ \citep{Jaffe03,svc08}.  The main
contributors to this nHz GWB signal are MBH systems of mass
$M>10^8\,\Msun$.  Current pulsar timing experiments limit the GWB
spectral amplitude to $\lesssim 7 \times 10^{-15}$, depending somewhat
on the value of~$\alpha$ \citep{h05,jhv+06,l+09}.  With longer spans of
data, PTA experiments become sensitive to lower GW frequencies, where
the expected signal is stronger.  The ongoing NANOGrav pulsar timing
program will achieve a GW sensitivity well into the predicted $h_c$
amplitude range in the next 3--5 years (Figures~\ref{fig:gwb} and
\ref{fig:sensitivity}).

As these limits improve, and ultimately progress to a detection, they
provide a new view of the history of MBH mergers throughout the
Universe.  Furthermore, measurements of the GWB spectral shape
near~10~nHz could distinguish between various models of MBH binary
formation \citep{svc08}.  LISA will be sensitive to the final MBH-MBH
coalescence events for systems with $M<10^7\,\Msun$.  Pulsar timing
arrays and LISA thus provide complementary views of these sources,
encompassing the full range of MBH masses.

\subsection{Gravitational Wave Bursts and Individual Sources}\label{sec:bursts}

Pulsar timing has already been used to constrain GW emission from
individual sources.  For example, a proposed MBH binary within the radio
galaxy 3C 66B \citep{Sudou03} was ruled out when its GW signature was
not detected in existing millisecond pulsar data sets \citep{jllw04}.

GW bursts (events lasting less than a few years) are also potentially
detectable.  Sources might include highly eccentric systems near
periapsis or the final inspiral of merging black holes \citep{l+09}.  GW
waveforms of such events would encode detailed information about the
burst source, such as the masses and spins of the inspiraling black
holes.

\subsection{Cosmic Strings and Exotica}\label{sec:strings}

Pulsar timing experiments may provide a unique window into particle
physics at the highest energy scales.  Cosmic strings, theorized
line-like topological defects, may form during phase transitions in the
early Universe, due to the rapid cooling that took place after the Big
Bang \citep{Kibble:1976sj}.  Recently it was shown that cosmic string
production is generic in supersymmetric grand unified theories
\citep{Jeannerot:2003qv}.  Furthermore, string theoretical cosmology
predicts the formation of so-called {\em cosmic superstrings}, different
from regular field theoretical cosmic strings \citep{Polchinski05}.

Cosmic strings and superstrings are expected to produce a stochastic GWB
analogous to the cosmic microwave background, as well as bursts of GWs
\citep{Damour01,damour2005,Siemens06}.  Because of their sensitivity at
very low frequencies, pulsar timing arrays place the best constraints on
viable cosmic string models~\citep{Siemens07}.  With increased
sensitivity, PTAs could detect cosmic (super)strings.  

Pulsar timing experiments could result in the detection of other
exotica.  Gravitational waves are a means for probing the fundamental
structure of the space-time of the Universe.  With less than 20\% of the
matter in the Universe emitting electromagnetic radiation, we are likely
to be surprised by what we ``see'' in GWs, the generation of which is
caused directly by the movement of mass, not the coupling to the
electromagnetic force.

\section{Key Advances For A Pulsar Gravitational Wave
Observatory}\label{sec:advance}

The sensitivity of a PTA is determined by the number and distribution of
the pulsars under observation, the cadence with which they are observed,
and the precision with which the pulse times of arrival are measured.

Pulsar timing precision is quantified by the root-mean-square (RMS)
residual pulse arrival time after a $\chi^2$ fit to a standard model of
pulsar rotation, binary motion, Earth motion, and interstellar
propagation effects.  The RMS residual for a given source is determined
by its flux density, characteristic pulse shape, emission stability,
rotation stability, the scintillation and scattering of its signal as it
traverses the interstellar medium, and the radio telescope equipment
used to observe it (telescope area, system temperature $T_{sys}$,
bandwidth, data acquisition instrumentation, and detection algorithms).
Currently there are several pulsars with RMS residuals approaching
100~ns, and roughly 20 more with residuals less than 1~$\mu$s.
\citet{Jenet05detect} showed that with 100-ns level timing on 20
pulsars, the stochastic GWB is detectable in $\sim$5 years.  Realizing
this goal will come from a two-pronged approach:  We must find
additional pulsars suitable for high-precision timing and also improve
the timing precision of known sources.

\citet{Demorest09} have recently suggested that published RMS pulsar
timing residuals over the past two decades show an exponential
improvement with time analogous to Moore's Law for computer processors
(Figure~\ref{fig:rms_vs_time}), improving by a factor of~2 every $\sim$3
years.  Several observational advances are required if we are to
continue this trend:

\begin{figure}[t]
\begin{center}
\includegraphics[width=0.9\textwidth]{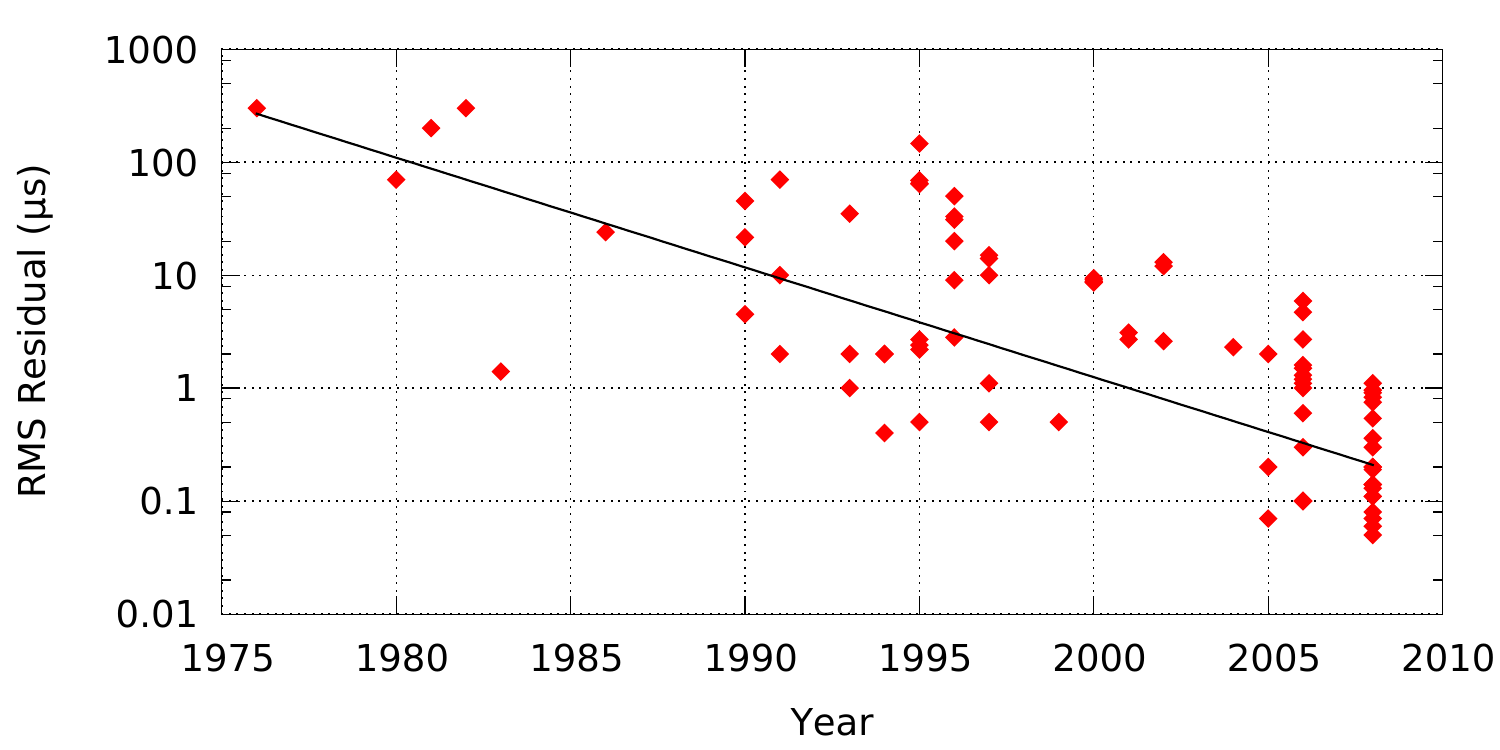}
\end{center}
\vspace*{-6ex}
\caption{\label{fig:rms_vs_time}%
Published RMS pulsar timing residuals versus time, showing exponential
improvement (Demorest \& Jenet, 2009) which positions us to detect GWs
within the next decade.
}
\vspace*{-1ex}
\end{figure}

\noindent {\bf A. Pulsar Surveys} The sensitivity of a PTA scales
directly with the number of pulsars in the array \citep{Jenet05detect}.
Given sufficient observing time, we expect to be make a detection of GWs
with currently known pulsars within 5 years.  However, to fully
characterize the gravitational waveforms and to maximize the scientific
return, more high-timing precision millisecond pulsars are needed.  It
is especially important to include more pulsars in directions widely
separated from the current set of objects.  Three current searches are
now turning up such objects: the PALFA L-band multibeam survey at
Arecibo, the GBT 350~MHz drift scan survey, and the new Parkes L-band
Digital Survey.  In the next decade, a new GBT low-frequency pulsar
survey would be particularly advantageous because it would increase the
number of Northern hemisphere pulsars.  This area of the sky is
currently under-represented in timing arrays.  As pulsar flux increases
at low radio frequencies (from $\sim$1~mJy at 1.4~GHz to $\sim$10~mJy at
400~MHz), this survey will identify many new pulsars.

\noindent {\bf B. Time on High Sensitivity Facilities} To achieve the
necessary precision on these faint objects we require multifrequency
observations on 100-m class or larger radio telescopes.  Four-frequency
observations are necessary in order to precisely fit for the
frequency-dependent dispersion caused by free electrons in the ISM,
which typically produces perturbations of $\sim$$\mu$s.  Table
\ref{tab:Arecibo_hours} illustrates the total worldwide telescope time
being devoted to pulsar timing with the goal of GW detection, weighted
by telescope sensitivity.  For reference, several future facilities are
listed: By 2011 five European telescopes will be combined into a phased
array, the Large European Array for Pulsars (LEAP), with an effective
area equivalent to a 200-m single dish.  Other future facilities are the
full 350-dish Allen Telescope Array (ATA-350), the Expanded Very Large
Array (EVLA) and the Square Kilometer Array (SKA).

\begin{table}[t]
\begin{center}
\vspace*{-1.75ex}
\begin{threeparttable}
\caption{\label{tab:Arecibo_hours} International PTA telescope time in
terms of a 100-m dish with $T_{sys} = 30$K.  
}
\begin{tabular}{lcccccc}
\hline
          & Diameter     & $\epsilon$\tnote{a} & $T_{sys}$ &  $\epsilon
          A / T_{sys}$ & Allocated   & 100-m equiv. \\
Telescope & (m)          &            & (K)      & (normalized)    & Time/mo (h) & time (h)  \\

\hline 
{\bf Current Projects} \\
Arecibo	     &   305      &  0.5  & 30 & 5.0  & 8            & 200   \\
Europe       &   $\sim$100&  0.7  & 30 & 0.7  & 125\tnote{b} &  60   \\
GBT          &   100      &  0.7  & 20 & 1.1  & 18           &  20   \\
Parkes       &    64      &  0.6  & 25 & 0.3  & 100          &  10   \\
\hline
{\bf Future Projects } \\
Europe-LEAP  &   200\tnote{c} &  0.7  & 30 & 3.0  & 24          & 220 \\
EVLA         &   130\tnote{c} &  0.5  & 30 & 0.9  & {\em TBD}   & --   \\
ATA-350      &   110\tnote{c} &  0.6  & 40 & 0.6  & {\em TBD}   & --    \\
SKA          &   750\tnote{c} &  0.6  & 35 &  30   & {\em TBD}  & --  \\
\hline
Total (Current)    &	   &         &    &       &	&   290\\ 
\hline
{\bf Requirements} \\
GW Detection\tnote{d}      &         &   &   &       & &   500\\
Advanced GW Study\tnote{e} &         &   &   &      & &  $>$1000 \\
\hline
\end{tabular}
\begin{tablenotes}
{\footnotesize
\item[a]{Includes the effects of reflector efficiency and partial illumination.}
\item[b]{This represents the combined observing time of four European
100-m class dishes.}
\item[c]{Equivalent single-dish diameter.}
\item[d]{20 pulsars with $\lesssim$100~ns RMS timing.}
\item[e]{$>$40 pulsars with $\lesssim$100~ns RMS timing.}
}
\end{tablenotes}
\end{threeparttable}
\vspace*{-5ex}
\end{center}
\end{table}

The table shows that world resources currently provide about 300 100-m
hours per month.  We estimate that at minimum, GW detection requires
$\sim$500 100-m hours per month, based on observing 20 pulsars every 2
weeks, for 3 hours at each of 4 radio frequencies in order to obtain
100-ns or better timing precision.  To fully characterize the GW sources
requires at least twice as many pulsars \citep{lee:gwpol}, and an effort
to upgrade receivers and backend instrumentation to handle $\sim$1~GHz
total bandwidth.  This process is underway at many existing radio
telescopes.  Clearly, to fulfill the goal of detecting and
characterizing the stochastic GWB, as well as continuous and burst
sources, we require more resources than are currently available.  The
table suggests that the ATA and EVLA could possibly provide the
additional time needed in the near future, and the SKA farther in the
future.

NANOGrav, the Parkes Pulsar Timing Array (PPTA) and the EPTA are in the
process of organizing themselves as the International Pulsar Timing
Array (IPTA) for the purpose of optimizing these international
resources.  In addition to sensitivity, a full optimiziation must
consider such additional factors as telescope sky coverage, frequency
agility, and backend instrumentation.

\noindent {\bf C. Algorithm Development} Algorithms for both radio pulse
detection and GW detection must be improved.  With increased telescope
sensitivity, pulse arrival times are increasingly susceptible to
systematic effects.  Among the algorithms to be developed are methods
for characterizing and compensating for the effects of the interstellar
medium \citep{Foster90, You07}; methods for effectively mitigating radio
frequency interference \citep{Stairs00}; and methods for fully utiziling
the available polarization information \citep{vanStraten06}.  Algorithms
for optimal extraction of the GW signal will be built upon the recent
advances of LIGO and LISA data analysis; some of this work has already
begun \citep{Jenet05detect, vanHaasteren08, Anholm08, l+09}.  

\section{Summary} Given sufficient resources, we expect to detect GWs
through the IPTA within the next five years.  We also expect to gain new
astrophysical insight on the detected sources and, for the first time,
characterize the universe in this completely new regime.  The
international effort is well on its way to achieving its goals.  With
sustained effort, and sufficient resources, this work is poised to offer
a new window into the Universe by 2020.

{\footnotesize
\setlength{\baselineskip}{0.75\baselineskip}
\begin{multicols}{2}
\bibliography{all_refs}

\begin{thebibliography}{}

\bibitem[{Anholm} et~al.(2008){Anholm}, {Ballmer}, {Creighton}, {Price}, and
  {Siemens}]{Anholm08}
{Anholm}, M., {Ballmer}, S., {Creighton}, J.~D.~E., {Price}, L.~R., \&
  {Siemens}, X. 2008, ArXiv e-prints

\bibitem[{Boroson} and {Lauer}(2009){Boroson} and {Lauer}]{boroson09}
{Boroson}, T.~A., \& {Lauer}, T.~R. 2009, ArXiv e-prints

\bibitem[{Comerford} et~al.(2008){Comerford}, {Gerke}, {Newman}, {Davis},
  {Yan}, {Cooper}, {Faber}, {Koo}, {Coil}, and {Rosario}]{cgn+08}
{Comerford}, J.~M., {Gerke}, B.~F., {Newman}, J.~A., {Davis}, M., {Yan}, R.,
  {Cooper}, M.~C., {Faber}, S.~M., {Koo}, D.~C., {Coil}, A.~L., \& {Rosario},
  D.~J. 2008, ArXiv e-prints

\bibitem[{Damour} and {Vilenkin}(2001){Damour} and {Vilenkin}]{Damour01}
{Damour}, T., \& {Vilenkin}, A. 2001, \prd, 64, 064008

\bibitem[{Damour} and {Vilenkin}(2005){Damour} and {Vilenkin}]{damour2005}
{Damour}, T., \& {Vilenkin}, A. 2005, \prd, 71, 063510

\bibitem[Demorest and Jenet(2009)Demorest and Jenet]{Demorest09}
Demorest, P., \& Jenet, F. 2009, in preparation

\bibitem[{Ferrarese} and {Merritt}(2000){Ferrarese} and {Merritt}]{Ferrarese00}
{Ferrarese}, L., \& {Merritt}, D. 2000, \apjl, 539, L9

\bibitem[{Foster} and {Backer}(1990){Foster} and {Backer}]{Foster90}
{Foster}, R.~S., \& {Backer}, D.~C. 1990, \apj, 361, 300

\bibitem[{Hobbs}(2005){Hobbs}]{h05}
{Hobbs}, G. 2005, {\em PASA}, 22, 179--183

\bibitem[{Hulse} and {Taylor}(1975){Hulse} and {Taylor}]{Hulse75}
{Hulse}, R.~A., \& {Taylor}, J.~H. 1975, \apjl, 195, L51

\bibitem[{Jaffe} and {Backer}(2003){Jaffe} and {Backer}]{Jaffe03}
{Jaffe}, A.~H., \& {Backer}, D.~C. 2003, \apj, 583, 616

\bibitem[{Jeannerot} et~al.(2003){Jeannerot}, {Rocher}, and
  {Sakellariadou}]{Jeannerot:2003qv}
{Jeannerot}, R., {Rocher}, J., \& {Sakellariadou}, M. 2003, \prd, 68, 103514

\bibitem[{Jenet} et~al.(2004){Jenet}, {Lommen}, {Larson}, and {Wen}]{jllw04}
{Jenet}, F.~A., {Lommen}, A., {Larson}, S.~L., \& {Wen}, L. 2004, \apj, 606,
  799--803

\bibitem[{Jenet} et~al.(2005){Jenet}, {Hobbs}, {Lee}, and
  {Manchester}]{Jenet05detect}
{Jenet}, F.~A., {Hobbs}, G.~B., {Lee}, K.~J., \& {Manchester}, R.~N. 2005,
  \apjl, 625, L123

\bibitem[{Jenet} et~al.(2006){Jenet}, {Hobbs}, {van Straten}, {Manchester},
  {Bailes}, {Verbiest}, {Edwards}, {Hotan}, {Sarkissian}, and {Ord}]{jhv+06}
{Jenet}, F.~A., {Hobbs}, G.~B., {van Straten}, W., {Manchester}, R.~N.,
  {Bailes}, M., {Verbiest}, J.~P.~W., {Edwards}, R.~T., {Hotan}, A.~W.,
  {Sarkissian}, J.~M., \& {Ord}, S.~M. 2006, \apj, 653, 1571--1576

\bibitem[{Kibble}(1976){Kibble}]{Kibble:1976sj}
{Kibble}, T.~W.~B. 1976, {\em J. Phys. A}, 9, 1387

\bibitem[{Lee} et~al.(2008){Lee}, {Jenet}, and {Price}]{lee:gwpol}
{Lee}, K.~J., {Jenet}, F.~A., \& {Price}, R.~H. 2008, \apj, 685, 1304--1319

\bibitem[Lommen et~al.(2009)Lommen, Finn, Hobbs, R, and Jenet]{l+09}
Lommen, A., Finn, L.~S., Hobbs, G., R, M., \& Jenet, F. 2009, in preparation

\bibitem[Phinney(2001)Phinney]{Phinney01}
Phinney, E. 2001, ArXiv e-prints

\bibitem[{Polchinski}(2005){Polchinski}]{Polchinski05}
{Polchinski}, J. 2005, {\em IJMP A}, 20, 3413--3415

\bibitem[{Rodriguez} et~al.(2006){Rodriguez}, {Taylor}, {Zavala}, {Peck},
  {Pollack}, and {Romani}]{rtzppr06}
{Rodriguez}, C., {Taylor}, G.~B., {Zavala}, R.~T., {Peck}, A.~B., {Pollack},
  L.~K., \& {Romani}, R.~W. 2006, \apj, 646, 49--60

\bibitem[{Sesana} et~al.(2008){Sesana}, {Vecchio}, and {Colacino}]{svc08}
{Sesana}, A., {Vecchio}, A., \& {Colacino}, C.~N. 2008, \mnras, 390, 192--209

\bibitem[{Siemens} et~al.(2006){Siemens}, {Creighton}, {Maor}, {Majumder},
  {Cannon}, and {Read}]{Siemens06}
{Siemens}, X., {Creighton}, J., {Maor}, I., {Majumder}, S.~R., {Cannon}, K., \&
  {Read}, J. 2006, \prd, 73, 105001

\bibitem[{Siemens} et~al.(2007){Siemens}, {Mandic}, and {Creighton}]{Siemens07}
{Siemens}, X., {Mandic}, V., \& {Creighton}, J. 2007, \prl, 98, 111101

\bibitem[{Stairs} et~al.(2000){Stairs}, {Splaver}, {Thorsett}, {Nice}, and
  {Taylor}]{Stairs00}
{Stairs}, I.~H., {Splaver}, E.~M., {Thorsett}, S.~E., {Nice}, D.~J., \&
  {Taylor}, J.~H. 2000, \mnras, 314, 459

\bibitem[{Sudou} et~al.(2003){Sudou}, {Iguchi}, {Murata}, and
  {Taniguchi}]{Sudou03}
{Sudou}, H., {Iguchi}, S., {Murata}, Y., \& {Taniguchi}, Y. 2003, Science, 300,
  1263

\bibitem[{van Haasteren} et~al.(2008){van Haasteren}, {Levin}, {McDonald}, and
  {Lu}]{vanHaasteren08}
{van Haasteren}, R., {Levin}, Y., {McDonald}, P., \& {Lu}, T. 2008, ArXiv
  e-prints

\bibitem[{van Straten}(2006){van Straten}]{vanStraten06}
{van Straten}, W. 2006, \apj, 642, 1004

\bibitem[{You} et~al.(2007){You}, {Hobbs}, {Coles}, {Manchester}, {Edwards},
  {Bailes}, {Sarkissian}, {Verbiest}, {van Straten}, {Hotan}, {Ord}, {Jenet},
  {Bhat}, and {Teoh}]{You07}
{You}, X.~P., {Hobbs}, G., {Coles}, W.~A., {Manchester}, R.~N., {Edwards}, R.,
  {Bailes}, M., {Sarkissian}, J., {Verbiest}, J.~P.~W., {van Straten}, W.,
  {Hotan}, A., {Ord}, S., {Jenet}, F., {Bhat}, N.~D.~R., \& {Teoh}, A. 2007,
  \mnras, 378, 493

\end{thebibliography}
\end{multicols}
\par}

\end{document}